# Growth Rates in Smith-Purcell FEL

K.V. Ivanyan*

M.V. Lomonosov Moscow State University, Moscow 119991, Russia

*k.ivanyan@yandex.com

**Abstract:** In the framework of the earlier derived dispersion equation, we study induced Smith–Purcell (SP) radiation of relativistic electron beam in the absence of a resonator. We offer a new method for calculation of coefficients for partial amplitudes in the case of rectangular grating. Using this method we calculate the growth rate of the SPFEL with a rectangular grating. The growth rate dependences on harmonic number, beam current, observation angle and relativistic factor are presented.

1. Overview

In [1] a single-particle classical theory of the Smith-Purcell free electron laser is described. The gain is optimized with respect to the grating period and the light incident angle.The optimal parameters of the device and the optimal gain are found. The latter is estimated to be of the order of several percent in thr IR frequency region. Here, only interactions of electrons with a light field reflected from a rippled surface is considered. Finite sizes of grating and of a light waist are taken into account.

In [2] it was found that the dispersion equation describing the induced SP instability is a quadratic equation for frequency; and the zero-order approximation for solution of the equation, which gives the SP spectrum of frequency, corresponds to the mirror boundary case, when the electron beam propagates above plane metal surface (mirror).

In [3] the simplest model of the magnetized infinitely thin electron beam is considered. The basic equations that describe the periodic solutions for a self-consistent system of a couple of Maxwell equations and equations for the medium are obtained.

In [4] for the grating, which has depth of grooves as a small parameter, the dispersion equation of the Smith-Purcell instability was obtained. It was found that the condition of the Thompson or the Raman regimes of excitation does not depend on beam current but depends on the height of the beam above grating surface. The growth rate of instability in both cases is proportional to the square root of the electron beam current.

There are numerous publications devoted to FELs on undulators and strophotrons [5-46 and references therein]. After the advent of Free Electron Lasers (FEL), it was suggested to use the Smith-Purcell effect to create a new type of FEL – the Smith-Purcell Free Electron Laser (SP FEL). There are also many publications devoted to SP FEL (see [47-51] and references therein).

In the present paper as an important example for the application of the obtained results of [2], we present a small signal gain of the SP FEL signal in the case with a rectangular grating.

## 2. Rectangular grating

In this section, as an important example of application of the obtained results of [2], we present the small signal gain of a SP FEL in the case of a rectangular grating.

In spite of the fact that our approach excludes the case of a rectangular grating on the face of it, the obtained equation (51) of Ref. [2] can describe that case. One of the ways is to calculate the coefficients in Eq. (51) of Ref. [2] with the method used in [47-51]. The solution of Maxwell equations is expressed as series of the cavity modes for the grooves space $y < h$. To sew the solutions for $y > h$ and for $y < h$ gives the coefficients for Eq. (51) of Ref. [2]. In our paper we offer another method. In order to find the coefficients $G_{np}$ we consider the grating with trapezoid form

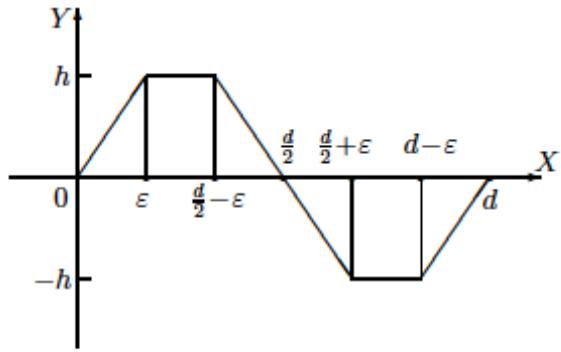

Fig.1. The function of f(x) given by Eq. (1).

$$f(x) = \begin{cases} \dfrac{h}{\varepsilon}x, & 0 < x < \varepsilon \\ h, & \varepsilon < x < \dfrac{d}{2} - \varepsilon \\ -\dfrac{h}{\varepsilon}x + \dfrac{hd}{2\varepsilon}, & \dfrac{d}{2} - \varepsilon < x < \dfrac{d}{2} + \varepsilon \\ -h, & \dfrac{d}{2} + \varepsilon < x < d - \varepsilon \\ \dfrac{h}{\varepsilon}x - \dfrac{hd}{\varepsilon}, & d - \varepsilon < x < d \end{cases} \quad (1)$$

Substituting the function (1) in formula Eq.(53) of Ref.[2] and then integrating we find

$$G_{nn}^{\pm} = 2(\pi - 2\eta)\cos(q_{ny}h) + \frac{4\eta}{q_{ny}h}\sin(q_{ny}h), \qquad (2)$$

$$\begin{aligned}G_{np}^{\pm} &= \frac{2\eta}{(n-p)\eta \pm q_{ny}h}\sin[(n-p)\eta \pm q_{ny}h] + \frac{2\eta(-1)^{n-p}}{(n-p)\eta \mp q_{ny}h}\sin[(n-p)\eta \mp q_{ny}h] \\ &- \frac{2}{n-p}\{\sin[(n-p)\eta \pm q_{ny}h] + (-1)^{n-p}\sin[(n-p)\eta \mp q_{ny}h]\}.\end{aligned} \qquad (3)$$

Here $\eta = \chi\varepsilon$ is a dimensionless length.

For the triangular form of grating $\eta = \pi/2$ and we get

$$G_{nn}^{\pm} = \frac{2\pi}{q_{ny}h}\sin(q_{ny}h), \qquad (4)$$

$$\begin{aligned}G_{np}^{\pm} &= \frac{\pi}{(n-p)\frac{\pi}{2} \pm q_{ny}h}\sin[(n-p)\frac{\pi}{2} \pm q_{ny}h] + \frac{\pi(-1)^{n-p}}{(n-p)\frac{\pi}{2} \mp q_{ny}h}\sin[(n-p)\frac{\pi}{2} \mp q_{ny}h] \\ &- \frac{2}{n-p}\{\sin[(n-p)\frac{\pi}{2} \pm q_{ny}h] + (-1)^{n-p}\sin[(n-p)\frac{\pi}{2} \mp q_{ny}h]\}.\end{aligned} \qquad (5)$$

For the rectangular profile of the grating $\eta = 0$ and we find

$$\begin{aligned}G_{nn}^{\pm} &= 2\pi\cos(q_{ny}h), \\ G_{np}^{\pm} &= \pm\frac{2}{n-p}\left[(-1)^{n-p} - 1\right]\sin(q_{ny}h).\end{aligned} \qquad (6)$$

With $|n|$ rising the coefficients $T_{np}$ increase. In order to remove the overfull data in numerical calculations we have to re-normalize the coefficients $T_{np}$:

$$\begin{aligned}T_{nn} &= \alpha_n, \\ T_{np} &= \frac{(-1)^{n-p} - 1}{\pi}\left(\frac{1}{n-p} + \chi\frac{q_{nx}}{q_{ny}^2}\right)\frac{\sin(q_{ny}h)}{\cos(q_{ny}h)}\left(1 + K(n)\left(1 + e^{2iq_{nb}}\right)\right).\end{aligned} \qquad (7)$$

for imaginary $q_{ny}$.

We can not make the analytic estimation of the upper limit for the absolute value of the term. Instead we made a numerical test. We took into account the (2N+1) surface modes, or in other words, we calculated the growth rate using (2N+1) x (2N+1) matrix. The results of calculations are shown in Fig.2 for the following parameters: mode of SP-radiation n=2, relativistic factor gamma_0=2, beam current $I_b$=2A/cm, period of the grating d=1cm, amplitude of the grating h=1cm and height of the beam b=1cm.

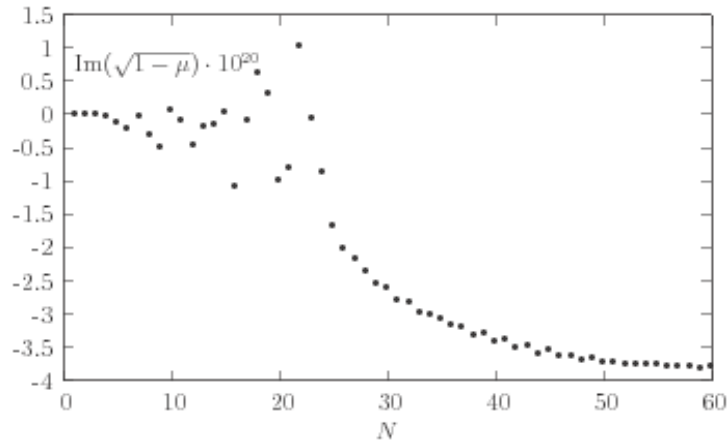

Fig.2. The dimensionless growth rate as a function of the matrix size (2N+1) x (2N+1) for theta =0.5rad. The calculation parameters are: mode of frequency n=2, gamma=2, beam current $I_b$ =2A/cm, period of grating d=1cm, amplitude of grating h=1cm, and height of the beam b=1cm.

We can see that the results depend weakly on N for N > 48. Therefore, in our calculation we have used N = 50.

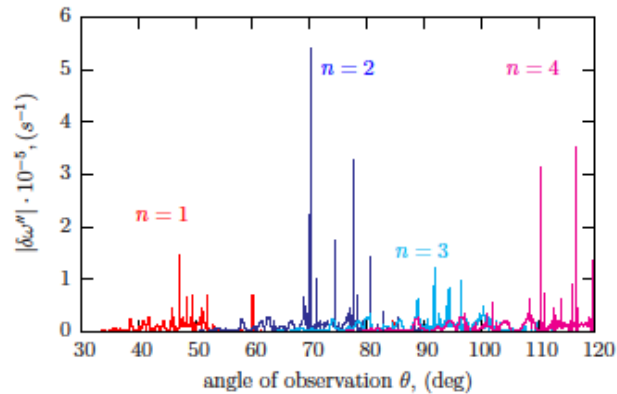

Fig.3. The growth rate as a function of the observation angle $\theta$ for different n modes of SP-radiation. The calculation parameters are: relativistic factor $\gamma = 4$, beam current $I_b$=10A/cm, period of grating d=1cm, amplitude of the grating h=1cm, and height of the beam b=1cm.

In Fig.3 the dependence of the growth rate on the observation angle $\theta$ is shown for the first four modes of the SP-spectrum for the following parameters of calculation: relativistic factor $\gamma = 4$, beam current $I_b$=10A/cm, period of grating d=1cm, amplitude of the grating h=1cm, and height of the beam b=1cm. The growth rate has got several high narrow peaks, the width of which is about 0.001~ rad. In real experiments this feature can be observed only if the real electron beams have narrow spread of electrons velocity. For the constant frequency $\omega_n$ the vagueness of the velocity $\delta\beta$ corresponds to the vagueness of the observation angle defined by the formula

$$\delta\theta = \frac{\delta\beta}{\beta^2 \sin\theta} \qquad (8)$$

For $\delta\beta \square 0.01 rad$ and $\beta \approx 1$, $\theta \approx 50 \deg.$ we find that $\delta\theta \square 0.01 rad$. This means that the peaks will be smoothed. So, the peaks of growth rate are realized when the velocity spread of the beam is smaller than $\delta\beta \square 0.01 rad$.

The maximal values of the growth rate for the first mode localize within the range of 37deg.< $\theta$ <60deg.. With the number of the SP spectrum mode increasing, the region, where the instability can excite, shifts to the large values of the observation angle $\theta$. For n=2 the maximal values of the growth rate are localized in the range of 55deg.< $\theta$ <100deg.. The maximum values of the growth rate for n=3 are in the interval of 70deg.< $\theta$ <110deg., while for n=4 are within the range of 90deg.< $\theta$ <120deg..

Fig.4. shows the growth rate as a function of the observation angle for parameters close to those of the Dartmouth experiment [52]: the Lorentz factor $\gamma = 1.068493$ (the corresponding electron energy E=35keV), and the distance between the beam and the top of grating $y_0 = 20 \mu m$ (the corresponding height of the beam b = 70 $\mu m$).

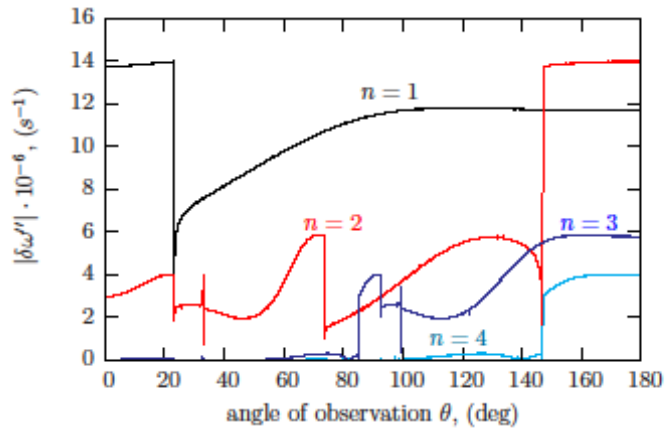

Fif.4. The growth rate as a function of the observation angle $\theta$ for different n modes of SP-radiation for parameters which are close to the parameters of the Dartmouth experiment [52].

The calculation parameters are: relativistic factor $\gamma = 1.068493$ (the corresponding electron energy E=35keV), beam current $I_b$=10A/cm, period of grating d=173 $\mu m$, amplitude of the grating h=50 $\mu m$, and height of the beam b=70 $\mu m$ (the corresponding gap between beam and top of grating $y_0 = 20 \mu m$).

Other parameters are beam current $I_b$=10A/cm, period of grating d=173, $y_0 = 20\mu m$, and amplitude of grating h=50, $y_0 = 20\mu m$. In this case the narrow peaks are absent. The first mode n=1 dominates in the range of 0deg.< $\theta$ <147deg., while the second mode n=2 dominates within the interval of 147deg.< $\theta$ <180deg.. So the mode with n=1 will excite first of all at $\theta$=90deg.. Our calculation well agrees with the results of the Dartmouth experiment, where only first mode was observed [52].

Fig.5. shows the dependence of the growth rate on the beam current $I_b$ for n=1 mode of SP-radiation and observation angle $\theta$ = 38.75 deg. The parameters of calculation are relativistic factor $\gamma$ =10, period of grating d=1 cm, amplitude of the grating h=1 cm, and height of the beam b=1cm. One can see that within wide range of current values the growth rate places under the law $\delta\omega'' \propto \sqrt{I_b}$.

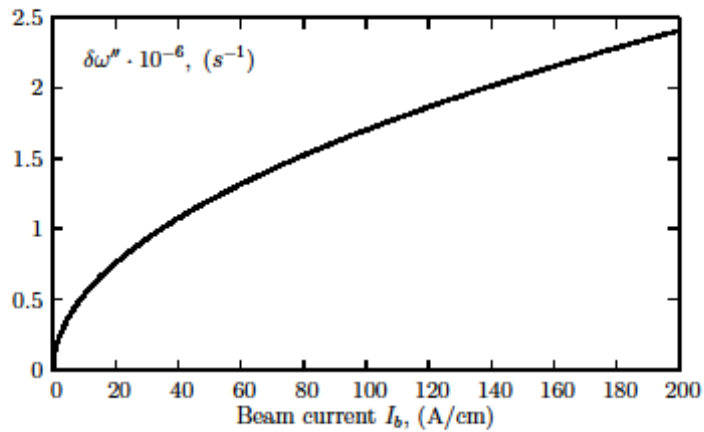

Fig.5. The growth rate as a function of the beam current $I_b$ for n=1 mode of SP-radiation and observation angle $\theta$=38.75deg. The calculation parameters are: relativistic factor $\gamma$=10, period of grating d=1cm, amplitude of the grating h=1cm, and height of the beam b=1cm.

Fig.6. shows the dependence of the growth rate on therelativistic factor gamma for the first three modes of of SP-spectrum for the following parameters of calculation: beam current $I_b$=10A/cm, period of grating d=1cm, amplitude of the grating h=1cm, and height of the beam b=1cm.

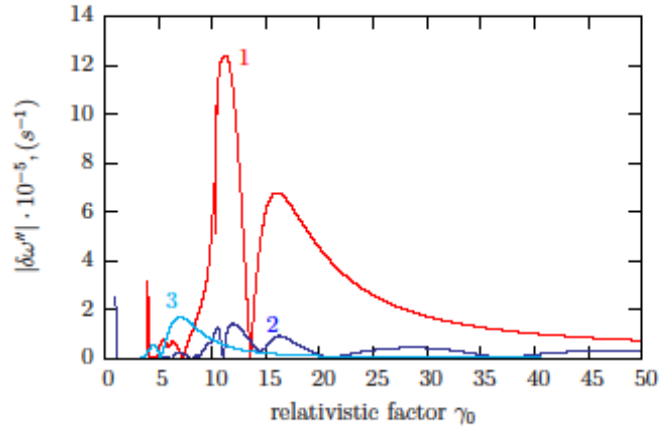

Fig.6. The growth rate as a function of the relativistic factor $\gamma_0$. (1) for n=1 mode of SP-radiation and observation angle $\theta$ =38.75deg, (2) for n=1 mode of SP-radiation and observation angle $\theta$ =32.83deg, (3) for mode n=3 of SP-radiation and observation angle $\theta$ = 80.21deg. The parameters of calculation are: beam current $I_b$=10A/cm, period of grating d=1cm, amplitude of the grating h=1cm, and height of the beam b=1cm.

The growth rate dependence on relativity marks a complicated form; function $\delta\omega''(\gamma)$ oscillates and decreases with the relativity rising. The maximum rate for the first mode corresponds to the high narrow peak. Our preliminary calculations show that the growth rate has of a complicated nature.

### 3. Conclusion

As an important example of the application of the obtained results in [2] the growth rate of SP FEL in the case with a rectangular grating was calculated. The calculated results are consistent with the experimental data obtained by Urata [52].

An analytical expression for the growth rate of SP instability has been derived. The gain dependence on the beam relativity factor and the beam height above the grating and the angle of observation has also been analyzed. We have examined various limits of the growth rate depending on the electron beam and grating parameters. Our calculations provided for rectangular grating are in good agreement with the results of the Dartmouth experiment, where only first mode of coherent emission by non-bunching beam was observed [52].

The author thanks D.N. Klochkov and K.B. Oganesyan for helpful discussion.**References**

1. Artemiev, A.I.; Fedorov, M.V.; Shapiro, E.A. Laser Phys., **4**, 1114 (1994).
2. A.I. Artemyev, M.V. Fedorov, A.S. Gevorkyan, N.Sh. Izmailyan, R.V. Karapetyan, A.A. Akopyan, K.B.Oganesyan, Yu.V.Rostovtsev, M.O.Scully, G.Kuritzki, J. Mod. Optics, **56**, 2148 (2009).
3. D.N. Klochkov, M.A. Kutlan, arxiv:1701.06159.
4. D.N. Klochkov, M.A. Kutlan, arxiv:1702.0242.
5. M.V. Fedorov. Atomic and Free Electrons in a Strong Light Field, Singapore, World Scientific, 1997.
6. Oganesyan, K.B. and Petrosyan, M.L., YerPHI-475(18) – 81, Yerevan, (1981).
7. Fedorov, M.V. and Oganesyan, K.B., IEEE J. Quant. Electr, vol. **QE-21**, p. 1059 (1985).
8. G.A. Amatuni, A.S. Gevorkyan, S.G. Gevorkian, A.A. Hakobyan, K.B. Oganesyan, V. A. Saakyan, and E.M. Sarkisyan, Laser Physics, **18** 608 (2008).
9. Zh.S. Gevorkian, K.B. Oganesyan Laser Physics Lett., **13**, 116002, (2016).
10. K.B. Oganesyan, J. Mod. Optics, **61,** 1398 (2014).
11. A.H. Gevorgyan , K.B. Oganesyan, Optics and Spectroscopy, **110**, 952 (2011).
12. Zaretsky, D.F., Nersesov, E.A., Oganesyan, K.B., and Fedorov, M.V., Sov. J. Quantum Electronics, **16**, 448 (1986).

    Zaretsky D F, Nersesov E A, Oganesyan K B and Fedorov M V., Kvantovaya Elektron. **13** 685 (1986).
13. Gevorgyan A.H., Oganesyan K.B., Karapetyan R.V., Rafaelyan M.S. Laser Physics Letters, **10**, 125802 (2013).
14. V.V. Arutyunyan, N. Sh. Izmailyan, K.B. Oganesyan, K.G. Petrosyan and Cin-Kun Hu, Laser Physics, **17**, 1073 (2007).
15. A.H. Gevorgyan , K.B. Oganesyan, J. of Contemporary Physics, **45**, 209 (2010).
16. K.B. Oganesyan, J. of Contemporary Physics**, 51**, 307 (2016).
17. E.A. Nersesov, K.B. Oganesyan, M.V. Fedorov, Zhurnal Tekhnicheskoi Fiziki, **56**, 2402 (1986).
18. Fedorov M.V., Nersesov E.A., Oganesyan K.B., Sov. Phys. JTP, **31,** 1437 (1986).
19. A.S. Gevorkyan, K.B. Oganesyan, Y.V. Rostovtsev, G. Kurizki, Laser Physics Lett., **12**, 076002, (2015).
20. K.B. Oganesyan, M.V. Fedorov, *Zhurnal Tekhnicheskoi Fiziki*, **57**, 2105 (1987).
21. D.N. Klochkov, K.B. Oganesyan, Y.V. Rostovtsev, G. Kurizki, Laser Physics Lett., **11**, 125001 (2014).
22. A.H. Gevorkyan, K.B. Oganesyan, E.M. Arutyunyan. S.O. Arutyunyan, Opt. Commun., **283**, 3707 (2010).
23. K.B. Oganesyan, Nucl. Instrum. Methods A **812,** 33 (2016).
24. A.H. Gevorgyan**,** M.Z. Harutyunyan, G.K. Matinyan, K.B. Oganesyan, Yu.V. Rostovtsev, G. Kurizki and M.O. Scully**,** Laser Physics Lett., **13,** 046002 (2016).
25. Petrosyan M.L., Gabrielyan L.A., Nazaryan Yu.R., Tovmasyan G.Kh., Oganesyan K.B., Laser Physics, **17**, 1077 (2007).
26. A.H. Gevorgyan, K.B.Oganesyan, E.M.Harutyunyan, S.O.Harutyunyan, Modern Phys. Lett. B, **25**, 1511 (2011); K.B. Oganesyan, J. Contemp. Phys., **50,** 123 (2015).